# An Uncommon Task: Participatory Design in Legal AI

FERNANDO DELGADO, Cornell University, USA
SOLON BAROCAS, Microsoft Research and Cornell University, USA
KAREN LEVY, Cornell University, USA

Despite growing calls for participation in AI design, there are to date few empirical studies of what these processes look like and how they can be structured for meaningful engagement with domain experts. In this paper, we examine a notable yet understudied AI design process in the legal domain that took place over a decade ago, the impact of which still informs legal automation efforts today. Specifically, we examine the design and evaluation activities that took place from 2006 to 2011 within the Text REtrieval Conference's (TREC) Legal Track, a computational research venue hosted by the National Institute of Standards and Technologies. The Legal Track of TREC is notable in the history of AI research and practice because it relied on a range of participatory approaches to facilitate the design and evaluation of new computational techniques—in this case, for automating attorney document review for civil litigation matters. Drawing on archival research and interviews with coordinators of the Legal Track of TREC, our analysis reveals how an interactive simulation methodology allowed computer scientists and lawyers to become co-designers and helped bridge the chasm between computational research and real-world, high-stakes litigation practice. In analyzing this case from the recent past, our aim is to empirically ground contemporary critiques of AI development and evaluation and the calls for greater participation as a means to address them.

## 1 INTRODUCTION

Contemporary critiques of artificial intelligence (AI) and machine learning (ML) systems emphasize the need for social scientific expertise and greater participation of domain stakeholders in AI system development [35, 47, 110, 139]. Expanding participation is often proposed as a strategy for identifying and mitigating risks brought about by the integration of AI systems into socially complex and high-stakes contexts. One rationale often articulated is that through increased cross-disciplinary collaboration, as well as greater inclusion of the people who will be interacting with the system, technologists will be able to better develop systems that are fit for purpose and real-world use.

Despite growing calls for participation in AI design, there are to date few empirical studies of what these processes look like and how they are structured for meaningful engagement with domain experts. In this paper, we examine a notable yet understudied AI design process in the legal domain that took place over a decade ago, the impact of which still informs legal automation efforts today. Specifically, we examine the design and evaluation activities that took place from 2006 to 2011 within the Text REtrieval Conference's (TREC) Legal Track, a computational research venue hosted by the National Institute of Standards and Technologies (NIST) [82]. The Legal Track of TREC is notable in the history of AI research and practice because it relied on a range of participatory approaches to facilitate the design and evaluation of new computational techniques—in this case, for automating attorney document review for civil litigation matters.

The purpose of the Legal Track was to explore how computational techniques might help to automate attorney document review as part of a broader civil discovery process through which adversarial parties cooperate on exchanging relevant factual information with each other. The aim of discovery is to equip both sides with a full view of the evidentiary basis of a case. As such, discovery serves as the critical backdrop for all subsequent phases of litigation, and is the phase in which outcomes of the majority of cases are determined [44]. The Legal Track created a unique collaborative environment in which, through an iterative and interactive simulation, litigators were provided visibility into the affordances and limitations of various data analytic approaches while computer scientists were provided visibility into the principles and mechanics informing civil discovery fact-finding—with both sides given the opportunity to develop a working understanding of the other's professional practices.

Given the very limited history of participation in AI design and evaluation, what accounts for this rare case of robust design collaboration? Drawing on archival research and interviews with various participants in the Legal Track, this paper aims to explain why participatory approaches took hold in this particular case of AI design and evaluation, how participation was operationalized in practice, and the extent to which these methods helped lawyers and computer scientists translate and resolve their respective concerns across disciplinary boundaries. In our analysis we find that the evaluation goals and design methods of Legal Track were in large part driven by domain practitioners rather than computer scientists, despite it being a computational research venue. Further, we find that by structuring design and evaluation activities through the format of an extended role-play simulation, Legal Track coordinators helped to neutralize the knowledge asymmetries between computer scientists and non-technical stakeholders. We also find that a key element of Legal Track's success as a site of ML translation into practice was its persistent embrace of real-world complexity and consistent rejection of task simplification in order to produce something immediately tractable.

This history is all the more remarkable since it was by no means a guaranteed outcome that AI would receive anything more than a lukewarm response from legal professionals. Legal practitioners have good reason to worry about failures in uncovering materials relevant to a case: attorney document review and fact-finding play a critical, often decisive role in the resolution of civil litigation matters [44]. The legal profession has also traditionally exhibited a reluctant stance toward integrating new forms of expertise into its procedures and workflows [45, 62]. It is thus particularly noteworthy that an AI-based approach for automating attorney review was able to achieve legitimacy as early as it did among an influential set of judges and litigators [64, 91, 112]. This paper thus also reflects on why participatory methods seemed to work so well in this case and what this experience might tell us more broadly about fostering effective participation in AI. In analyzing this case from the recent past, our aim is to empirically ground contemporary AI critique and analysis, especially with respect to calls for greater participation in AI design. Our findings provide valuable insights into the role that participation might play more broadly in the translation of AI research into practice, as well as the development of AI workflows that stand up to legal and other forms of normative scrutiny.

We proceed as follows. In the next section, we provide an overview of the role discovery plays in U.S. civil litigation and introduce Technology Assisted Review (TAR), an AI-driven approach to classifying documents for review [51]. In Section 3, we review scholarship on legal automation, participatory design, and participation in AI design, and situate our work in the CSCW and broader literature. Section 4 details our research site, materials, and methodological considerations. The next three sections of the paper report our findings, organized thematically. Section 5 details the major historical events that lead to the establishment of the Legal Track, the key figures who orchestrated that process, and the unique way that they decided to structure the annual convening. Section 6 delves into the participatory nature of the interactive task in the Legal Track, uncovering a range of mechanisms that facilitated effective participation by legal professionals. Section 7 describes how key statistical concepts from computer science were mapped onto values and outcomes underpinning civil discovery practice—and how efforts to automate manual workflows drew new scrutiny to the effectiveness of manual practices. We show how this helped set the stage for the broader litigation community of practice to develop a working understanding of quantitative evaluation metrics in ML and their value in applying ML to civil discovery practice. We turn in Section 8 to a broader discussion of the lessons that can be drawn from this account, as well as certain limitations. Section 9 concludes.

## 2 BACKGROUND

### 2.1 Discovery in Civil Litigation

Civil discovery is a procedure whereby parties to a lawsuit are able to obtain evidence from another party [69]. The goal of discovery is to help support specific party claims, defenses, and assessments of damages in order to prepare for a trial, or to help litigants decide whether to settle a case, based on the available evidence [124]. Discovery, in this light, can be conceptualized as a truth-seeking procedure through which evidence is compelled into view and the claims of parties are judged based on the available evidence. The process requires adversarial parties to cooperate on exchanging relevant factual information with each other so that both sides can develop a full view of the evidentiary basis of a case. In this vein, discovery's value necessarily depends on how comprehensively the process is conducted—without broad discovery, the ability to uncover the facts required in making a careful and just assessment of the merits of the claims is constrained. We might think of modern American discovery practice as a sort of "data-driven" mode of adjudication.

Discovery processes are particularly important—and particularly challenging—in the context of complex litigation. Taking the largest securities fraud class action recovery in history as an example, in 2002, the energy firm Enron was accused of defrauding shareholders through false statements and deceptive accounting practices [61]. The plaintiffs' claims were substantiated by internal documentation and emails unearthed during the discovery phase of the litigation. Faced with incontrovertible evidence of the scheme, Enron's company directors, as well as co-defendant banks, auditors, and law firms, contributed to an overall $7.2 billion settlement fund—the largest securities class action recovery to date. As with any other discovery effort, central to the work of discovery in this case was the identification of relevant documents in Enron's vast internal databases.

The problem has only become more severe over time: when large corporations or government actors are involved in civil litigation, and their petabytes of internal data are subject to discovery, parties must shoulder a high cost in responding in a timely and accurate manner to the scores of requests that they may receive. These obligations have traditionally been discharged through attorneys manually reviewing each document to assess its responsiveness, a laborious process that requires significant investments of money and time. In its 2012 empirical survey and analysis of discovery costs, the RAND Corporation's Institution of Civil Justice urged the legal community to "move beyond its current reliance on [manual] eyes-on review" in order to avoid an already labor-intensive process from becoming intractable and risking the timely resolution of high-stakes matters—in other words, to shift away from heavily manual attorney review flows by adopting some form of automation [88].

## 2.2 Technology Assisted Review (TAR)

To address these challenges, civil litigants in the United States increasingly rely on TAR. The vast majority of TAR implementations make use of a supervised ML framework [51]. In a supervised ML framework, the algorithm learns how to distinguish between responsive and non-responsive documents based on the presence or absence of combinations of features including words, phrases, conceptual clusters, punctuation, and metadata.

The transformation of litigation review from a painstaking manual process to a sophisticated algorithm-driven methodology took place over a relatively short period of time. As we will discuss below, TAR was introduced to a handful of litigators in an experimental research setting in 2008 [82], and was first deployed on a civil case only shortly thereafter, in 2012 [37]. By 2015, a vocal and influential set of jurists was actively advocating for TAR's use in cases involving significant volumes of documents for attorney review [e.g., 97]. Since then, TAR has spread across the U.S. civil court system, and its use is governed by a substantial set of procedural rules (including the Federal Rules of Civil Procedure [23]) and case law derived from the scores of cases in which the use of TAR has been adversarially negotiated. Moreover, legal professional associations such as the American Bar Association [e.g., 12], the Sedona Conference [e.g., 109], and the Duke Bolch Judicial Institute [e.g., 19] regularly publish primers and hold meetings regarding TAR. For any lawyer working in civil litigation, it has now become a matter of core professional competence to be versed in the affordances of TAR. They are expected to have sufficient knowledge to be able to advise clients on whether and how to leverage TAR and subsequently defend these choices to opposing parties and the presiding judge [23].

## 3 RELATED WORK

### 3.1 Legal Automation

The prospect of automating legal workflows has existed at least since Vannevar Bush's 1945 description of the speculative memex device that would give way to "wholly new forms of encyclopedias [...] with a mesh of associative trails," and, for an attorney in particular, provide "at his touch the associated opinions and decisions of his whole experience, and of the experience of friends and authorities" [25]. This then-futuristic and optimistic account contrasts with much of the contemporary scholarship on legal AI, including TAR. Much of the current work on legal AI reflects legal practitioner-oriented concerns regarding negative impacts on attorneys, including: the reduction of learning opportunities afforded to early career litigators as substantial swaths of document review and analysis work become automated [65], the risk of undermining client representation as new technical personnel—not subject to lawyers' ethical obligations—weigh in on legal decisions [42, 94], and more generally the loss of legal professional agency as more litigation work becomes the purview of non-lawyers [58, 65, 89, 94, 95]. Additionally, legal commentators have expressed concerns regarding the conflict between the due process norm of participation in the U.S. civil justice system and the costly expertise required to implement and audit TAR successfully [42]. These concerns, based on reasonable perceived threats posed by the introduction of algorithmic techniques into attorney workflows, echo critiques of decision automation systems in administrative law [27, 30, 33, 34], as well as ongoing concerns related to automated risk assessment tools used in the criminal justice system [3, 36].

In the HCI and CSCW literature, legal automation is largely absent from the discussion. To our knowledge, there are no works that specifically examine TAR despite the various interesting challenges presented by introducing AI into an already complex professional workflow, as well as TAR's growing importance in the U.S. civil justice system. Expanding the review of existing HCI and CSCW scholarship to include work on legal automation more broadly, two interesting early CSCW pieces come into view [16, 123]. Both of these works focus on a much earlier wave of legal discovery automation in which paper documents were being electronically scanned, stored, and then bibliographically coded (as a pre-step for making them indexable and searchable for attorney review). Unlike TAR, in which the considered judgment of trained lawyers is targeted for automation, this early wave of discovery automation targeted work performed by litigation support staff.

A prescient aspect of these early CSCW works addressing legal automation was their experimentation with participatory design (PD) techniques to better understand the discovery domain. They made use of collaborative prototyping to make sense of a complex workflow in which there were "overwhelming logistical requirements" and a "multiplicity of actors involved" with various "unfolding interests" [16]. Their work foreshadows contemporary calls for incremental design interventions [137] in how they refocus their efforts away from developing new systems whole cloth and toward "embedding bits of automation" that would "relieve the tedium" for the litigation support staff while ensuring their "interactive control over the process necessary for interpretation and the exercise of judgment" [124].

### 3.2 Participatory Design

More broadly, PD in HCI and CSCW has historically aimed to push back on a perceived over-reliance on abstraction and formalism in technology development. Early PD researchers and practitioners were concerned that incumbent design methods modeled workplaces from a distance with only idealized views of work practices, thus failing to capture important tacit understandings and knowledge held by workers [125]. They sought to counteract this elision by leveraging techniques like "cooperative prototyping" [18] that served to produce "co-constructed artifacts" between designers and work practitioners [16, 124]. Through these collaborative design efforts, a shared space could be created that serves as a region owned by neither technologist and designer—or what Muller and Druin call a "third space" [81]—in which relationship building and mutual learning could take place.

PD also responds to concerns regarding the asymmetry of power and knowledge often found in technology development in which computer scientists and designers hold relative advantage over most other stakeholders. By using non-formal modes of communication such as mock-ups and prototypes—rather than highly specialized technical language—asymmetries could be neutralized, at least in theory [21, 22]. For PD adherents, designers risk falling into a participation trap in which stakeholders are invited to the table yet not able to provide meaningful input if there isn't sufficient focus at first on establishing a shared language to collaborate with non-designers. In other words, for PD, process is key in order to engender substantive change in how decisions are made and who makes them in the design of technology; "you can't just add users and stir" [81].

### 3.3 Participation in AI Design

A growing body of literature at the intersection of the AI, CSCW, and HCI research communities is now critically examining how PD and related critical design methods can positively impact the development of AI systems on the ground. For example, a number of workshops have been convened to explore the role PD (or participation more broadly) can play in the design of AI systems [73, 136, 139], and several articles have been published advocating for increased use of PD methods in AI [13, 72]. These calls complement the design and policy recommendations formulated by critical AI researchers concerned with the lack of multi-stakeholder involvement in AI system design, development, and governance [35, 47, 110, 138]. They also echo earlier efforts by HCI researchers who championed the use of PD methods for earlier rule-based AI systems [76, 115, 122].

The current wave of PD research in AI includes the development of new systems for social media content moderation [28, 53, 140], as well as decision-support tools in human services [24, 78] and hospital care [111]. Additionally, a growing set of HCI evaluation studies in domains including child welfare [104], prostate cancer diagnosis [26], and diabetic retinopathy [14] have been undertaken with domain experts to better understand the organizational contexts for AI use in order to inform downstream PD interventions.

A shared feature across these design efforts is their focus on modeling the human expertise and labor central to the overall maintenance of these systems. This centering of workers and non-technical professionals in technical system design can be traced back to early Scandinavian collaborative design efforts for technologies whose introduction into corporate settings would have an impact on working conditions and workers' skill sets [119]. In a contemporary context focused on algorithmic disruption of workflows, these researchers adopt a participatory process as a way to prevent "de-professionalization" via automation and to explore how human discretion can be designed into algorithmic decision-making workflows [78]. PD in a contemporary AI context also serves to promote new organizational relationships and ways of communicating that strengthen the internal capability to take ownership of algorithmic systems and repair them when failures arise [111].

Techniques from PD are also used by researchers to induce acceptance or bolster comfort in a system that has already largely been designed. Some examples here include the growing interest in Explainable AI (XAI) to solicit user input on and community approval of human interface features for an already existing AI system [70, 79, 80]. In public service contexts, techniques originating from PD practice have been leveraged "to raise comfort levels among affected communities" in the development and deployment stages of AI systems [20], as well as for the system roll-out and community outreach process [138].

Drawing upon and extending this prior work, the current paper offers an empirical account of the design process that helped to establish ML as a crucial component of litigation practice. In doing so, this case study can serve as a useful exemplar to chart a path forward for AI design methodology more broadly.

## 4 SITE & METHODS

### 4.1 TREC Legal Track

The Legal Track of TREC ran for six years between 2006 and 2011 [82]. The larger TREC conference itself has a rich history, given its relation to the broader development of information retrieval (IR) as a field, its historical connection to the Defense Advanced Research Projects

Agency (DARPA), and the various fields it has served in fostering state-of-the-art (SOTA) text processing and search technologies since 1992 [101, 134]. TREC is a modern example of a classic IR-style evaluation framework—known as the Cranfield method—in which various researchers across government, industry, and academia perform experiments on shared test collections to compare effectiveness across different retrieval techniques [31, 32, 54, 55, 134].

TREC can also be categorized as an example of a technical evaluation and benchmarking venue making use of the Common Task Framework (CTF) in which a set of enrolled competitors test out various methods on publicly available training datasets, while being evaluated by an impartial scoring referee [71]. CTF has been recently historicized as the 'secret sauce' behind the many advances we see today in ML. Its key ingredient is competition driven by leaderboard tracking, resulting in an extreme focus on incremental gains in accuracy [17, 39]. Yet TREC differs substantially in tone from what has been described as a 'CTF Lifestyle' in which teams participate in a SOTA-chasing, winner-take-all competition, with cash prizes to boot [40].

While quantified metrics are central to evaluation in TREC, TREC coordinators describe their brand of common task exercises as rooted more broadly in "collections, comparisons, and communities," notably framing their event as a conference rather than as a competition [84]. How TREC's stated focus on communities was actually operationalized in the context of Legal Track is of key interest for this study. While the initial rounds of Legal Track were aligned with the traditional Cranfield method and CTF, the task design and evaluation approach adopted by Legal Track coordinators evolved significantly after the first year—in large part to respond to the unique complexity of the legal discovery community's work. The bulk of these changes served to expand Legal Track's contribution beyond data set development and quantitative benchmarking to include more participatory and qualitative forms of research and assessment.

### 4.2 Methodological Approach

We examine Legal Track's participatory approach to stakeholder involvement as an important historical precedent in AI design practice. We frame this research as an empirical case study [1, 92, 131] meant to shed light on the early design efforts supporting the integration of AI/ML techniques into legal decision-making workflows. As such, Legal Track provides us with an illustrative case study [118, 120] that can help us understand how participatory processes are best suited to inform translation of AI research into real-world practice [67]. Further, this work is inspired by a growing set of work in the CSCW and HCI literature regarding the need for a greater focus on design history [38, 117, 118].

The data used in our analysis consist of both archival reports and semi-structured interviews. We retrieved primary source material from the official Legal Track documentation repository [82]. For each of the six years the conference ran, the full set of the following sets of documents were qualitatively reviewed, coded, and analyzed [127]: instructions, guidelines, and FAQ documentation ("Protocols") drafted by coordinators and distributed to participating teams, final reports ("Overviews") published by coordinators, project post-mortem memos ("Lessons Learned") authored by domain experts, and research publications ("Proceedings Papers") submitted by teams. In total, we analyzed 76 official documents related to Legal Track. We supplemented this archival analysis with 26 publications including research articles and legal commentary where Legal Track was discussed in depth. In total, we included 102 documents in our analysis.

In addition, to further contextualize our archival analysis, seven semi-structured interviews were conducted, transcribed, coded, and analyzed over a two-month period (June to July 2021). The interviewees included the majority of Legal Track coordinators across its multi-year run and all Legal Track coordinators who served in that role across at least two iterations (and thus who had a role in not only coordinating the track within one year, but also a role in devising at least one future round based on the previous years' experience). All interviews were approximately one hour in length, and began with basic questions regarding how interviewees were introduced to the Legal Track, what role they had as coordinators, how they felt Legal Track was similar as well as different from other TREC tracks they participated in, and how they would describe the interaction between computer scientists and litigators on Legal Track across the years it ran.

In tandem, the corpus of Legal Track primary source documents and the set of semi-structured interviews with Legal Track coordinators provide us with a rich view into the task protocols, rationales for design, results, analysis, concerns, and recommendations coming from the participants involved in Legal Track.

### 4.3 Terminology

The actors involved in Legal Track never explicitly used terms such as "participatory design" or "co-design." We use these terms because they map to useful frameworks in the HCI and CSCW literature regarding the type of cross-disciplinary and stakeholder interaction that is of core interest to this study. The actors involved in Legal Track do not describe their work as "design" as much as "evaluation"; and yet, while TREC is not traditionally considered an HCI or CSCW venue, we interpret many of their research questions, protocol designs, and modes of reflective analysis as centrally relevant to design questions concerning human-machine interaction and collaboration.

The actors involved also do not generally describe Legal Track activities under an "artificial intelligence" rubric, but more specifically as an "information retrieval" effort (though they published in AI journals and conferences, and retrospectively do frame TAR as an AI approach). This in part follows from the fact that Legal Track (2006-2011) took place before an explosive growth in AI research and commercialization brought about by advances in deep learning. The supervised machine learning workflow that TAR is based on emerges from an IR/text analytics tradition that considerably predates deep learning. Also worth noting along these lines is that "machine learning" as it is referenced in the Legal Track archive is not treated as a separate field or specialization (as often occurs today), but rather one of many computational techniques that IR researchers and engineers may leverage.

While for our purposes there is no ambiguity or controversy in designating TAR as an AI technology, it is important to note these semantic shifts that have occurred between now and then in order to carefully map between historical description and contemporary analysis. It is also relevant to call out from a historical perspective how Legal Track uniquely straddles a period before the ascendance of deep learning in which symbolic and statistical approaches to AI are in conversation and competition with one another. Legal Track thus spans different waves of AI research, consists of a broad range of design, evaluation, and translation activities, and sits at the intersection of academic research, commercialization efforts, and government innovation policy.

Within this trajectory, we fix our empirical focus specifically on how a supervised ML approach to document categorization was developed and evaluated by a research community of practice (i.e., IR researchers) and a legal community of practice (i.e., civil discovery practitioners), informing the successful translation of ML research into real-world litigation. Importantly, we do not intend this analysis as an appraisal of contemporary TAR practice. Digging into the current state of TAR practice is a separate project requiring different data sets and sociolegal methods to properly conduct (and one which will benefit from this foundational historical work). Our focus here is on excavating TAR's design history—not to uphold it uncritically as an exemplar for best practices in AI design, but to examine it as a previously undocumented precedent of AI design that can provide us with interesting lessons for AI practice going forward.

### 4.4 Positionality: Actor as Analyst

The first author of this article was drawn to TAR as a research topic in reflecting on his previous experience as a legal technologist at an electronic discovery firm. He began his industry career in 2005 and for over a decade played several roles including data scientist, team lead, engagement manager, and organizational head. In becoming acquainted in more recent years with how AI/ML was being integrated into other domains such as policing, criminal justice, and human resources, he became curious about the unique role Legal Track had in the introduction of AI/ML expertise into the civil litigation domain. The first author participated marginally in the efforts of the team from his firm in the 2008 and 2009 rounds. He was therefore aware of the venue, yet played neither a central role in a team nor any role on the Legal Track organizational committee across Legal Track's life span.

## 5 BIG TOBACCO ORIGINS

### 5.1 Deluge of White House Emails

We begin our account of the Legal Track with a Racketeer Influenced and Corrupt Organizations Act (RICO) case filed by the Department of Justice during the Clinton Administration. United States v. Philip Morris USA was one of the first cases in which it was broadly recognized that traditional manual attorney review and keyword workflows were failing to provide efficient and comprehensive discovery. By the time it ended in 2006, despite both sides incurring enormous costs, neither side was able to assess whether it had delivered a truly comprehensive production in response to discovery requests [74]. The U.S. government itself struggled to respond to Philip Morris's requests (delivered in a 1,726 paragraph document), later disclosing that it took 25 archivists and lawyers six months to review approximately 200,000 White House emails that had been flagged by relevant keywords—leaving no time to do any due-diligence-oriented sampling or searching of the remaining 19.8 million records that had not been flagged by rudimentary keyword search [6, 7].

Reflecting on the process, Jason Baron, the Director of Litigation at the National Archives and Records Administration (NARA), was convinced that the era of "manual-review intensive" discovery was over [4]. For Baron, and other litigators in similar positions across public interest and corporate litigation, the exponential growth in electronically stored information across corporations and government agencies had brought about an undeniable "stress point in litigation" that risked indeterminately delaying the resolution of large matters, and which, for cases small and large alike, was causing discovery costs to often exceed the amount in the underlying dispute [90]. Baron voiced a broader consensus forming among litigators when he pushed for examining a "family of computer technology employing new types of search methods and techniques beyond the use of mere keywords" to help address the structural issues the community was facing with the arrival of the data deluge [90].

### 5.2 Practitioner-Driven Prospecting

The failure of keywords to weed out a high number of false positives—that is, documents that were erroneously flagged as responsive to discovery requests, when they were not in fact relevant—was a central problem Baron noted early on [4]. Yet, an even deeper concern held by Baron at that time was what he would later describe as the "dark matter" in discovery—those documents that were relevant, but were not captured by keywords and therefore not queued for manual review [6]. Baron and other litigators were in search of, as he put it, a "better machine" than keyword search to both better weed out false positives and also better address potential false negatives. He also understood that the legal discovery community needed an empirical mechanism to evaluate candidate approaches and benchmark their ability to help litigators honor their discovery obligations [4].

Concerned that manual review and keyword-based approaches were failing civil discovery, Baron reached out to Ellen Voorhees, an expert on text analytics in charge of the TREC program, to see if there was a way to structure an evaluation exercise around the needs of the legal community [6]. In parallel, he approached Douglas Oard, a leading text analytics researcher and colleague of Baron's at the University of Maryland, as well as Richard Braman, the Executive Director of the Sedona Conference, a nonprofit legal think tank dedicated to finding solutions to complex litigation matters [96]. Baron and Oard drafted a proposal that would eventually be approved by Voorhees and the other members of the TREC steering committee to establish the Legal Track.

Beyond the specific actors and institutions involved, what is notable here is how—rather than the computer scientists actively prospecting civil discovery as a domain in need of computational ordering [96, 114]—it was the non-technologist who was speculating, lobbying, and mobilizing for technological interventions in the legal domain. While the computer scientists involved in TREC were very responsive to and engaged with the problems civil discovery was facing once they were brought into the picture, there is no indication that any of them assumed they were sure to have a silver-bullet solution, nor does it appear they looked upon the legal domain opportunistically as a site to further advance IR as an "engine of progress" or "universal(izing) science" [96, 114]. In fact, the computer scientists specifically involved in the establishment and coordination of Legal Track were very keen on squashing "audacious claims" made by marketers regarding IR tools, and took care to correct and temper what they saw as legal experts' imprecise understandings and overly optimistic appraisal of advanced text analytic approaches, including topic modeling and other non-Boolean techniques then informally lumped under the label of "concept search" [86].

## 5.3 Devising a Common Task for the Legal Domain

Once Legal Track was scheduled to kick off in 2006, the next step was to establish a shared dataset to which all teams would have access. David Lewis, a veteran member of the TREC steering committee, played a critical role in procuring and making available Legal Track's first data set— seven million documents produced for discovery in lawsuits filed by the Attorneys General of several states against seven major tobacco organizations [8]. Using a publicly available data set related to Big Tobacco litigation was only fitting given that it was a DOJ case against Phillip Morris that initiated Baron's search for better text retrieval and classification tools in the first place. The "Master Settlement Agreement" (MSA) database, as it was referred to, consisted of a wide array of document types including letters, memoranda, budgets, reports, minutes, transcripts, agendas, scientific articles, and some email [8]. This heterogeneity was essential for evaluating different computational approaches, as real-world litigation discovery often consists of diverse document types. Later in 2009, to address the sparsity of email in the MSA database, Legal Track coordinators added a second dataset from the In re Enron Corporation Securities Litigation class action. This second dataset was comprised of approximately 850,000 documents, consisting of email messages and attachments [57].

In addition to shared data, coordinators needed to articulate realistic categorization tasks to make for a meaningful and challenging evaluation exercise. To facilitate this, a group of lawyers associated with the Sedona Conference created five fictional complaints, covering claims involving improper campaign contributions, deceptive advertising, insider trading, antitrust behavior, and product liability [8]. These complaints were then further mapped to specific requests for document production which became the basis for subject matter topics that teams would be tasked to computationally model. For example, for the deceptive advertising complaint, teams were asked to identify all documents related to Big Tobacco's practice of placing tobacco products onto television shows and movies, including the strategy, negotiation, and execution of these product placements. To keep repeat teams on their toes, litigators developed new fictional complaints each year expanding the range of claims to include medical malpractice, international trade violations, patent infringement, and securities fraud [57, 87, 129]. As a further categorization task example, for the fictional securities fraud complaint, teams were asked to identify all documents related to energy contracts agreed to by the defendant including the estimates, forecasts, and analyses that informed the negotiation and servicing of those contracts.

## 6 INTERACTIVE SIMULATION

### 6.1 Computer Scientists Wrestle with Relevance

Given the inherent ambiguity and variability of human language, identifying all—or even the majority of—relevant documents in a large unstructured dataset is an extremely difficult challenge [105]. In internet search—the most common object of IR study at this time—a user could often be satisfied with a handful of relevant results. Discovery review, however, represented an example of an exhaustive search exercise, in which a majority of all relevant items should be delivered [99]—a new and unfamiliar challenge for most new Legal Track participants. Teams participating in Legal Track were thus challenged to reconceptualize their working understanding of what adequate performance looked like in an IR task setting.

Further, the relevance criteria in discovery reviews were not of the same order as those that would have been familiar to teams participating in other TREC tracks. As one team relayed, "the information need is often so complex that it is hard to describe" in traditional query language [93]. Other teams struggled with the "fundamental vagueness" of document requests [43]. Indeed, in order to be successful, Legal Track teams needed to move beyond thinking of Legal Track topics as standalone specifications to be mapped directly to traditional IR queries, and instead to engage with them as underspecified document requests to be interpreted within a broader context of litigation strategy and legal procedure.

In addition to vagueness, another challenge was that relevance shifted over time. The traditional construction of relevance in IR was relatively static—but the type of relevance being sought in more complex search contexts consisted of a more emergent target that matured based on how a user "progresses through various stages of a task" and that was also sensitive to the "cognitive state" of the user (i.e., the set of information to which they have access and how they have synthesized that information) [92]. Applying this insight to a discovery context, participating Legal Track teams had to reckon with the fact that what made a document relevant to a topic in the eyes of an attorney could change over the duration of a review as new information was introduced via the various ongoing and iterative parallel processes of document analysis, motion practice, and legal strategizing [10, 11].

### 6.2 The Interactive Task

To begin addressing some of this complexity, coordinators of the 2007 Legal Track introduced an "Interactive Challenge Task." This was not TREC's first task designated as "interactive," but it did represent a new way of thinking about how to bring humans into the loop in IR research. While much was learned through previous interactive tasks, no new model or methodology for IR research or evaluation was synthesized from these efforts [34, 93], leaving Legal Track coordinators to chart their own course for how to leverage human input in IR research and evaluation.

Legal Track's Interactive Task aimed to "provide a more 'end-to-end' representation of the task of e-discovery in the real world, incorporating aspects of the challenge of e-discovery" [10] and did so in part by heavily recruiting input from legal community stakeholders to help with "objectively modeling the e-discovery review process for the purpose of evaluating the efficacy of a wide range of search methodologies" [96]. In other words, Legal Track's conceptualization and operationalization of interactivity relied heavily on multiple forms of community participation to support the research and evaluation effort. To "members of the bench"—that is, judges—it was presented in "Open Letters" and "Calls for Participation" jointly authored by The Sedona Conference and NIST as an independent research venue in which "both long established and emerging search methods for document review" were being evaluated and benchmarked [107, 108]. To industry stakeholders, the Interactive Task was presented as "an unprecedented opportunity to be at the forefront of an important movement to evaluate document review processes and create industry best practice standards" [96]. And perhaps most importantly, it was presented to the legal profession as a whole as a venue which aimed to model "how real lawyers would go about propounding discovery in civil litigation," focusing on how to "model as accurately as possible the real-world conditions in which companies and law firms, and the e-discovery firms they engage, must meet their document-retrieval objectives and obligations" [10].

## 6.3 Litigation Role-Play

While experienced legal professionals played several key roles in Legal Track from the outset—not only providing the original impetus for its development, but devising hypothetical complaints and requests, serving as document assessors, and playing the role of expert searchers to help identify algorithmic misses [8]—it wasn't until 2008 that coordinators defined a special role of Topic Authority (TA) for experienced litigators in the Interactive Task. Specifically, a TA was responsible for serving "as an authoritative source of information for participating teams seeking to develop definitions of relevance [and] serv[ing] as final arbiters of the samples reviewed to measure participating teams' effectiveness" [10]. For Legal Track coordinators, introducing a TA into the mix who interfaced directly with teams throughout the development process was a significant change to the task structure that more closely modeled how an "attorney employs the products or services of an e-discovery firm [...] with the goal of efficiently applying that conception of relevance across the full document population implicated by the matter" [10].

This interorganizational framing of the Interactive Task simulation, and the TA role in particular, was borne out of a critical insight gained by Legal Track coordinators in earlier rounds: it was the senior litigator who, at the end of the day, was responsible for ensuring an accurate and complete response to discovery requests and as such was the person whose conception of relevance mattered the most [11]. As a complement to this, it was the job of a team participating in Legal Track (as stand-in for an outsourced technology vendor) to "replicate, across the document population, one conception of relevance—that of the legal architect who has hired" them [10].

Overall, the TA role was designed with three chief responsibilities in mind. First, the TA provided topic clarification, "act[ing] as a resource to which teams can turn in order to clarify the scope and intent of a target topic" [11]. Second, the TA provided authoritative guidance about document relevance to participants, based on the coordinators' reasoning that "[i]n order to be able to obtain valid measures of effectiveness, it is essential that the samples reviewed for purposes of evaluation be assessed in accordance with the TA's conception of relevance" [11]. Finally, TAs served as final adjudicators for the task.

As such, TAs played more than merely a "human in the loop" supervisory role over the algorithmic systems devised by teams. These domain experts didn't merely train and evaluate the automated systems being developed; rather, the process was structured to instantiate a norm for having litigators train and evaluate the actual technologists developing the system.

### 6.4 Shifting Power and Knowledge Dynamics

Allowing the Legal Track TAs to not only help evaluate results at the end, but also opine on what they observed regarding each team's workflow, positioned Legal Track TAs as crucial experts. TAs could prescribe a host of remedies to bolster the awareness of the computer scientists participating in TREC regarding the interpretational complexity inherent to attorney review [15]. This positioning put domain experts on a more even playing field with the computer scientists than they might have been in a more traditional technical R&D setting. And this, in turn, resulted in interesting feedback dynamics emerging, including the TAs chiding teams for their "failure [...] to develop a systematic and deliberate plan for defining relevance, as opposed to applying an ad hoc approach" [15]. In addition to suggesting a need for tighter management in systems' development, they also pointed to problems with Legal Track teams' ability to undertake requirements gathering: "Some teams did not appear to grasp how to formulate questions that would have helped them to clarify the topic's scope or to arrive at a common conception of relevance" [15].

TAs instructed teams to place greater emphasis on their requirements and design approaches, including exploration of the "document collection early in the process," generation of "nuanced questions" connected to "exemplar documents" to be posed to the TAs and coordinators to inform system design and classification targets, and the overall establishment of an "iterative process that allowed for course correction" [15]. TAs also sometimes called out overly solutionist orientations by some teams: Though TAs "expected the teams to simulate an actual service provider focused on finding all of the documents the TA would need to identify to respond to the request for production[, some teams] were focused on understanding how a single technique could contribute to meeting e-discovery needs" [15].

Despite these concerns, there was consensus across coordinators, TAs, and teams alike that the more time teams spent with the TAs, the better [85]. Interaction with TAs was perceived by the teams as key to understanding how to model relevance effectively and specifically helpful in isolating which "document features such as document types and specific fields" could be best utilized [135]. Teams particularly found orientation-style meetings held by TAs on topic scope very helpful in correcting their "initial understanding of topic requests" and additionally appreciated when document requests were further contextualized by broader discussion of the mock complaint [93]. In contrast, some teams regretted not interacting more with TAs, and in their reports indicated that their failure to actively engage with TAs contributed to their low recall on the task [116, 126].

## 7 COUNTING ON MEASUREMENT

### 7.1 Mapping Numbers to Values

As is standard practice in IR, recall for Legal Track was calculated by looking at the percent of relevant documents identified by each team in relation to the total number of relevant documents existing in the corpus. Precision, in turn, was calculated by looking at the percent of correctly identified relevant documents in the result set from each team [49]. From the beginning, Legal Track coordinators understood that, in terms of the precision/recall trade-off (since gaining ground in one is in frequent tension with gaining or even maintaining ground in the other), recall should always trump precision in a discovery context [8, 9]. Recall was considered crucial for civil discovery review "since in real settings many requests for production state that 'all' such evidence is to be produced" [87]. Precision was also deemed important in order "to reduce unnecessary review costs" [87].

The use of an F measure, integrating both precision and recall numbers, was adopted in Legal Track "as a way of simultaneously reflecting the importance of recall (for exhaustiveness) and precision (for timeliness and affordability)" [87]. Mapped as such, precision and recall paired together were construed as proxies for tracking alignment between technological use and desired normative outcomes. This mapping allowed stated trade-off rationales regarding system design and performance to be interpretable across Legal Track coordinators and participants, computer scientists, and litigators alike.

The notion of recall played a critical role in contextualizing for Legal Track coordinators what made discovery review so distinct from other contexts. Generally, for the Legal Track coordinators who were IR researchers, it was a humbling experience to see many of the IR techniques leveraged in other domains (such as internet search) not quickly rise to the challenge of addressing discovery review needs [86]. They pointed to the "high recall task" nature of discovery review as the key challenge that needed to be addressed, distinguishing it from internet search or recommender tasks in which user satisfaction could often be achieved by finding and prioritizing just a few relevant items (and thus where precision plays a significantly more important role than recall). In civil discovery, the goal was to find all the relevant items as a class, and neither incumbent methods of retrieval nor evaluation of that retrieval were fit to this purpose [86].

In other TREC tracks, the evaluation process relies on relevance pooling: focusing formal assessment on only those documents across the benchmarking corpus that at least one team has identified as relevant [55, 98]. This pooling method allows assessment to focus on a constrained set of documents, on the assumption that, taken together, teams will uncover the vast majority of the relevant documents. Yet, for a high recall task, that assumption is rarely sound. It is very likely that all participating teams will miss relevant documents that are for one reason or another hard to computationally model. To address this gap, Legal Track coordinators specified a new protocol to examine the documents that all teams had classified as irrelevant, in the event that all teams were perhaps overlooking certain relevant documents. It would of course have been infeasible to review the whole "discard pile," so the Legal Track coordinators devised a sampling protocol to review a targeted volume of documents from this set. Doing so allowed the coordinators to develop more reliable estimates of recall.

Legal Track coordinators, participating teams, and TAs were not the only stakeholders tracking and debating metrics for automated approaches for attorney discovery review. By 2011, judicial commentary on Legal Track was honing in specifically on TAR. Judge Andrew Peck of the Southern District of New York stated that if the use of TAR "is challenged in a case before me, I will want to know what was done and why that produced defensible results. I may be less interested in the science behind the 'black box' of the vendor's software than in whether it produced responsive documents with reasonably high recall and high precision" [91]. What is especially worth noting here is not only the invocation of IR metrics in the judicial discourse on technological acceptance, but also the central role Judge Peck ascribes to them in addressing questions of procedural defensibility.

### 7.2 Debunking Human Review as the Gold Standard

In addition to interacting with the participating teams, the TA also had the responsibility of performing quality control on the first-pass review of documents conducted by law students and professional attorney document reviewers. The TA thus ensured that there was sufficient and adequately labeled data for the purposes of task evaluation. This particular role unexpectedly provided the TAs increased visibility into a more fundamental problem plaguing manual attorney review: human inconsistency and error. In their role as adjudicator at the end of the 2009 round, TAs reported being jolted by "flagrant errors" in the labeling of training and validation data provided by the document reviewers during the adjudication phase of results. The TAs subsequently commented on the need to reckon with the possibility "that human review may be more flawed than the legal professional currently understands and acknowledges" [15]. This stance echoed previous practitioner commentary regarding the "myth that manual review by humans of large amounts of information is as accurate and complete as possible – perhaps even perfect – and constitutes the gold standard by which all searches should be measured" [5].

This finding was subsequently integrated into the motivations commonly offered for TAR in the technical literature [50, 99], and triggered a broader reframing of the problem facing civil discovery to one not just of document volume, but also a more fundamental issue with human error in consistently applying interpretation guidelines in document review. With this reformulation, the discussion of automating discovery began to parallel other debates about the automation of white-collar work: the problem is not only information glut, but also faulty human decision-making [29]. This line of thinking culminated in what would subsequently be referred to in the TAR community as "the JOLT article" (referring to the Richmond Journal of Law and Technology) [49]. In this article—whose first author was both a Legal Track TA and coordinator—TAR (described as an "interplay of humans and computers") is introduced for the first time to a larger legal community. In the JOLT article, results from Legal Tracks 2008 and 2009 are presented to argue that TAR could lead to results that are more effective—and not just more efficient—than manual attorney review. This article became the established go-to reference when litigants required scientific support for their use of TAR, and is still cited by judges and litigators in orders and motions pertaining to TAR.

## 8 DISCUSSION

Legal Track enabled a deeply collaborative, participatory co-design process that pushed well beyond mere consultation with domain experts. The reasons for this are numerous and each offers more general lessons for the future of AI development and for making AI design more inclusive and participatory.

First, Legal Track's structure as a role-play simulation—with an interactive and iterative workflow—helped to neutralize the knowledge asymmetries between computer scientists and non-technical stakeholders. Enabling litigators, as domain practitioners, to participate in (and help to design) this simulation required significant scaffolding by Legal Track coordinators. This involved multiple steps, including getting litigators involved with the task set-up and training, and giving them a platform for evaluation in which they were given the opportunity to air extensive feedback. Through their performance of the TA role, litigators occupied a unique perch from which they could learn about computational tools and practice, and from which they could observe and analyze discovery review in a new light. In order to promote more collaboration and knowledge-sharing across disciplinary boundaries, Legal Track coordinators intentionally honed the roles of both litigators and teams through increasingly interactive task protocols. This process was demystifying in at least two respects: it gave litigators insight into the capabilities of the technology, and gave computer scientists insight into the complexities of the legal domain. This bilateralism was crucial for ensuring that the design process was successful and genuinely collaborative. As a more general lesson for AI design processes, Legal Track suggests the promise of processes in which expertise flows in multiple directions, rather than being extractive or hierarchical.

Second, Legal Track seemed to be successful because it embraced the complexity of the real-world task AI was being leveraged to address. Computer scientists involved in Legal Track did not respond to the many obstacles they encountered in applying their toolset to civil discovery by simplifying the problem formulation or reducing the scope of analysis to something immediately tractable. Rather, year after year, coordinators created or updated their datasets, commissioned hypothetical complaints and documents requests, and devised new administrative and evaluation protocols, all to better approximate the complexity of the real world and to reflect the lessons learned from previous iterations. Their aim was to develop systems that could actually address the scalability issues civil discovery was confronting in the face of exponential growth in digital data. The computer scientists coordinating Legal Track were even skeptical from the outset about how useful advanced analytics would be for discovery review. And when teams seemed more interested in testing out their particular tool than addressing the automation challenges of discovery review, Legal Track gave space for this type of solutionism to be critiqued as misaligned with the goals of the evaluation exercise.

Third, it is noteworthy that the idea for a Legal Track was actually devised by a representative of the legal domain and was not the brainchild of a computer scientist prospecting a new domain in which to apply their methods [cf. 96, 114]. This helps to explain why the litigators involved in Legal Track had such central roles in structuring classification tasks, evaluating teams' methods and outputs, and educating technologists on the complexity of their practice domain. While this unique participatory approach to AI design emerged in part as a response to the complexity of civil discovery review, it was also partially enabled by the fact that the computer scientists and litigators interacting with each other were often of equal professional status, coming from fields of similarly high prestige. It is perhaps this unique pairing of litigators soliciting computer scientists, and computer scientists consistently deferring to legal professional expertise, that helped to create a dynamic in which the substantive concerns arising from the legal domain held in balance the technical goals of computer science.

In assessing how to best generalize from Legal Track as an AI design precedent, we should note that Legal Track entailed a great deal of time and resources across a significant number of organizations over several years. This would be the case for other common task exercises, yet one leveraging the type of iterative and participatory techniques that the Interactive Task did necessarily requires an exceptional level of commitment and attention. Without proper incentives and resources to sustain this level of participation across the various actors involved, this type of design approach will be out of reach to most communities. Indeed, it is worth noting which stakeholders did not take part in the Legal Track. In addition to having document review lawyers and experienced litigators perform key roles in the overall evaluation process, a more dedicated focus on public interest concerns could have helped identify a broader set of actors from civil society who also hold a critical stake in civil litigation as potential plaintiffs or defendants. A primary question to address in thinking about applying participatory simulations to other contexts is thus how to provide the necessary support to allow a wider range of stakeholders to participate in the process. It also highlights the challenge of incentivizing researchers to commit to a multi-year research effort—and particularly one in which the design and evaluation techniques primarily evolve to address domain-specific problems, rather than necessarily hewing to broader trends in academic research and publishing in their home disciplines.

In addition to these questions of resources and incentives, another key aspect to address in thinking through the applicability of participatory simulations is the extent to which the application domain benefits from pre-existing rules of conduct and standards of care. The computer scientists involved in Legal Track did not need to come up with new rules for what it meant to properly conduct document review. The techniques for automation that they proposed were evaluated based on already established and codified rules of professional duty and care, as well as rules for evidence and ascertaining relevance. Further, the computer scientists' statistical approaches for evaluating accuracy were analyzed and mapped by legal stakeholders to their own procedural ways of thinking through pragmatic trade-offs when faced with the challenge of ensuring both comprehensive discovery and speedy justice. Without established institutional rules or norms to help clarify what sorts of AI-enabled results could be considered successful, it is less clear that a meaningful simulation could be designed or conducted, or how its results could be interpreted in a way that carried weight across a broader field of practice.

Taking these caveats into consideration, we can still view Legal Track as a useful example of how computer scientists can engage with domains in ways that are more bilateral and less extractive or transactional. It also showcases how participatory technology development can move forward without relying on increasing abstraction and reduction at each step. And it offers a rich view into how wrestling with the complexity of a domain can productively unsettle computational practice.

## 9 CONCLUSION

Through Legal Track, experienced litigators were provided a central role that enabled them to learn about the affordances and limitations of various data analytic approaches from leading computer scientists; conversely, computer scientists learned about the principles and mechanics informing legal fact-finding and attorney review from experienced litigators. In large part thanks to this structured collaboration, ML-driven methods tailored to the requirements of litigation practice were developed to meet the needs of the community. This process fostered a cohort of cross-disciplinary experts who learned how to effectively translate and design across computational and legal disciplinary boundaries, and who were instrumental to the early and principled integration of AI into U.S. civil litigation practice.

Our findings are especially relevant to contemporary calls for greater participation in AI design as a mechanism for integrating the concerns of a wider range of stakeholders, fostering greater accountability, and limiting downstream harms. Our findings demonstrate possibilities for thinking about participation from a relationship-building standpoint in which domain stakeholders play active and ongoing roles in driving problem formulation and determining evaluation criteria. The experience of the Legal Track of TREC highlights how such an approach helped bridge the chasm between AI academic research and real-world, high-stakes practice, and how PD processes—when structured to engender deep engagement from domain experts—may help to provide a basis for designing tools that meaningfully address community needs.

**ACKNOWLEDGMENTS**

This work was supported by the Russell Sage Foundation (Grant # 1908-17713) and the John D. and Catherine T. MacArthur Foundation. Any opinions expressed are those of the authors alone and should not be construed as representing the opinions of the Foundations.

51:20                                                                                     Fernando Delgado, Solon Barocas, & Karen Levy